\documentclass[12pt]{article}
\pdfoutput=1

\usepackage{graphicx}
\usepackage{amsmath, amssymb}
\usepackage{slashed}
\usepackage{color}
\usepackage{hyperref}

\hypersetup{colorlinks,bookmarksopen,bookmarksnumbered,citecolor=blue,
linkcolor=black,pdfstartview=FitH,urlcolor=blue}
  
\newcommand{\Slash}[1]{{\ooalign{\hfil/\hfil\crcr$#1$}}}

\begin{document}

\begin{titlepage}

\def\thefootnote{\fnsymbol{footnote}}

\begin{center}

\hfill KANAZAWA-18-06\\
\hfill TUM-HEP/1168/18 \\
\hfill December, 2018

\vspace{0.5cm} {\Large\bf Probing pseudo Nambu-Goldstone boson dark
  matter at loop level}

\vspace{1cm}
{\large Koji Ishiwata}$^{\it (a)}$ and 
{\large Takashi Toma}$^{\it (b)}$ 

\vspace{1cm}

{\it $^{(a)}${Institute for Theoretical Physics, Kanazawa University,
    Kanazawa 920-1192, Japan}}

\vspace{0.2cm}

{\it $^{(b)}${Physik-Department T30d, Technische Universit\"at M\"unchen, James-Franck-Stra\ss{}e, D-85748 Garching, Germany}}

\vspace{1cm}

\abstract{In the standard model extended by a single complex scalar
  boson with a softly broken global $U(1)$ symmetry, a pseudo
  Nambu-Goldstone boson becomes a candidate for dark matter. In this
  paper, we discuss the direct detection of the pseudo Nambu-Goldstone
  boson dark matter. Since the tree-level amplitude for dark
  matter--nucleon scattering vanishes, higher order quantum
  corrections for the amplitude should be taken into account. We
  perform the calculation at the next-to-leading order in QCD in a
  systematic manner.}

\end{center}
\end{titlepage}

\renewcommand{\theequation}{\thesection.\arabic{equation}}
\renewcommand{\thepage}{\arabic{page}}
\setcounter{page}{1}
\renewcommand{\thefootnote}{\#\arabic{footnote}}
\setcounter{footnote}{0}

\section{Introduction}

Although it is well-known that non-baryonic dark matter (DM) exists in
the universe, nature of dark matter is still a mystery of the universe
except the fact that its relic abundance occupies about $26\%$ of the
total energy density of the
universe~\cite{Aghanim:2018eyx,Ade:2015xua}.  A prominent candidate
for DM is a stable and non-relativistic particle that has the weak
scale interactions. In this case, the annihilation rate determining
its relic abundance is closely correlated with scattering rates with
the Standard Model (SM) particles and production rates at collider
experiments through the crossing symmetry.

Direct detection experiments of cold dark matter explores scattering
events between DM and nuclei. So far, any viable signal of DM has not
been found even in recent ton-scale detector experiments, which leads
to the bounds on the interaction between DM and nucleon.  The most
stringent upper bound on the spin independent (SI) elastic scattering
cross section between DM and nucleon is given by the XENON1T
Collaboration~\cite{Aprile:2018dbl}, which is, for example,
$4.1\times10^{-47}~\mathrm{cm}^2$ at the DM mass of $30~\mathrm{GeV}$
at $90\%$ confidence level.

The current null results of direct signal of DM motivate us to
consider a framework where the interactions between DM and nucleon are
suppressed in non-relativistic limit.  One of the ideas is to consider
a pseudo Nambu-Goldstone DM~\cite{Gross:2017dan} or a pseudo scalar
portal fermionic DM~\cite{Berlin:2015wwa, Arcadi:2017wqi,
  Abe:2018emu}. In the former case all the interactions between DM and
SM particles are described by derivative couplings at the tree level,
while in the latter case the DM spinor product for elastic scattering
is proportional to the DM velocity in non-relativistic limit.  As a
result, the elastic scattering amplitude is necessarily suppressed by
the small DM velocity at tree level, thus these DM candidates can
naturally be consistent with the strong constraints from the current
direct detection experiments.

In this paper we consider a pseudo Nambu-Goldstone DM proposed in
ref.~\cite{Gross:2017dan}, and study the possibility to detect it
directly.  As stated above, the elastic scattering amplitude for
direct detection is suppressed at the tree level in this model, and it
can be exactly zero in non-relativistic limit. However, the scattering
amplitude is expected to be finite at the loop level.  We will perform
the calculation at one-loop level for non-QCD part and at the
next-to-leading order level in QCD based on the formalism given in
ref.~\cite{Hisano:2015rsa} where the scattering processes with gluon
in nucleon, which are sometimes missed in the literature, are
systematically taken into account. Although the next-to-leading order
calculation in QCD is not necessary for a rough evaluation, it gives
${\cal O}(10\%)$ corrections in the amplitude and the theoretical
uncertainty regarding perturbative QCD calculation is reduced
significantly. We explore a parameter space consistent with the
observed DM relic abundance, the SM Higgs boson decay, and the
perturbative unitarity bound. We also compare the predicted elastic
scattering cross section with the sensitivity of the future direct
detection experiment DARWIN~\cite{Aalbers:2016jon}.

\section{The Model}
\label{sec:2}

We consider the SM augmented by a complex scalar field $S$ with a
softly broken global $U(1)$ symmetry. The model is invariant under the
transformation $S\to e^{i\alpha}S$ with a real constant $\alpha$
except for the soft breaking term. The scalar potential of the model is
given by
\begin{equation}
 \mathcal{V}=-\frac{\mu_{\rm H}^2}{2}|H|^2-\frac{\mu_{\rm S}^2}{2}|S|^2
 +\frac{\lambda_{\rm H}}{2}|H|^4+\lambda_{\rm HS}|H|^2|S|^2
 +\frac{\lambda_{\rm S}}{2}|S|^4
  -\left(\frac{\mu_{\rm S}^{\prime2}}{4}S^2+\mathrm{H.c.}\right)\,,
\end{equation}
where $H$ is the $SU(2)_L$ Higgs doublet which couples to the SM
particles.  The last term corresponds to the soft breaking term of the
global $U(1)$ symmetry. Due to the tachyonic mass terms for the scalar
fields, both $H$ and $S$ acquire vacuum expectation values (VEVs) in a
wide range of the parameter space, which is the situation we are
interested in. Then these fields are expanded around the vacuum as
\begin{equation}
 H=
  \left(
   \begin{array}{c}
    G^+\\
    \frac{1}{\sqrt{2}}(v+h+iG^0)
   \end{array}
  \right)\,,\qquad
  S=\frac{v_s+s+i\chi}{\sqrt{2}}\,,
\end{equation}
where $G^+$ and $G^0$ are the Nambu-Goldstone bosons associated with
the electroweak symmetry breaking, $v~(\simeq 246~{\rm GeV})$ and
$v_s$ are the VEVs for $H$ and $S$, respectively.  $h$ and $s$ are the
CP even scalar fields while $\chi$ is the CP odd scalar field which is
the would-be Nambu-Goldstone boson. Due to the soft breaking term of
the global $U(1)$ symmetry, non-zero mass for $\chi$ arises. Even
after the symmetry breaking, a $\mathbb{Z}_2$ symmetry remains, which
stabilizes $\chi$ and makes it a candidate for DM.

Due to the symmetry breaking, the CP even states $h$ and $s$ mix via
the Higgs portal coupling $\lambda_{\rm HS}$. We will derive the mass
eigenstates $h_1$ and $h_2$ at one-loop level. As mentioned in
Introduction, $\chi$--$q$ scattering amplitude vanishes at the tree
level. Therefore, one-loop corrections are necessary for studying
direct detection of $\chi$ DM. To this end, we execute the calculation
following ref.~\cite{Brignole:1992uf}. In the literature, the inverse
propagators for the scalars are calculated diagramatically. Then the
mass matrices for the CP odd and even sectors are defined by taking
zero external momenta. That corresponds to the one obtained from the
effective potential.  Then the mass eigenstates are given by
diagonalizing the mass matrices. In the following calculation we adopt
Landau gauge and $\overline{\rm MS}$ renormalization scheme as in the
literature. All couplings, scalar fields, and VEVs are renormalized
values. We will see that the renormalization scale dependence is
cancelled in the amplitude as expected.

The mass matrices for the CP even and CP odd sectors are given by
\begin{align}
&M^2_{\rm even} = \left(
    \begin{array}{cc}
      \lambda_{\rm H} v^2+\frac{T_h}{v} & \lambda_{\rm HS}vv_s\\
      \lambda_{\rm HS}vv_s & \lambda_{\rm S} v_s^2+\frac{T_s}{v_s}
    \end{array}\right)
    \equiv\left(
    \begin{array}{cc}
      M^2_{hh} & M^2_{hs} \\
      M^2_{hs} & M^2_{ss} 
    \end{array}  \right) \,,\\ 
  &M^2_{\rm odd}= \left(
    \begin{array}{cc}
      \frac{T_h}{v} & 0\\
       0 & \mu_{\rm S}^{\prime 2}+ \frac{T_s}{v_s}
    \end{array}\right)
    \equiv\left(
    \begin{array}{cc}
      M^2_{G^0G^0} & 0 \\
      0 & M^2_{\chi\chi} 
    \end{array}\right) \,,
\end{align}
where $T_h$ and $T_s$ are renormalized tadpoles for $h$ and $s$, which satisfy
the stationary conditions:
\begin{align}
  &\mu_{\rm H}^2 -\lambda_{\rm H} v^2 -\lambda_{\rm HS}v_s^2
  +\frac{2T_h}{v}=0 \,, \\
  &\mu_{\rm S}^2-\lambda_{\rm S}v_s^2 -\lambda_{\rm HS}v^2+\mu_{\rm S}^{\prime 2}
  +\frac{2T_s}{v_s}=0.
\end{align}
Then, the inverse propagators are given as
\begin{align}
  \Gamma_{ij}(p^2)
  &=\delta_{ij} p^2 -M_{ij}^2+\Pi_{ij}(p^2) \nonumber \\
  &=\delta_{ij} p^2 -\bar{M}_{ij}^2+\Delta \Pi_{ij}^2(p^2) \,,
  \label{eq:Gamma_ij}
\end{align}
where indices $i,j$ represent $h,s$ (or $h_1,h_2$), $G^0,\chi$, and
$\Pi_{ij}(p^2)$ correspond to the renormalized self-energies with the
external lines $i$ and $j$.  The concrete expressions for
$\Pi_{ij}(p^2)$ are collected in Appendix.  Here we have introduced
the quantities $\Delta \Pi_{ij}(p^2)$ defined by
\begin{equation}
  \Delta \Pi_{ij}(p^2)\equiv \Pi_{ij}(p^2)-\Pi_{ij}(0)  \,,
\end{equation}
and the definition of $\bar{M}^2_{ij}$ in Eq.~(\ref{eq:Gamma_ij})
follows accordingly. Note that $\bar{M}^2_{ij}$
correspond to the mass matrix derived from the effective potential,
{\it i.e.}, zero external momenta.  Since the mass matrix $M_{\rm
  odd}^2$ is diagonal, the physical (pole) masses for the CP odd
fields are simply given by $\Gamma_{G^0G^0}(0)=0$ and
$\Gamma_{\chi\chi}(m_{\chi}^2)=0$ where $m_\chi$ is the pole mass of
$\chi$, {\it i.e.},
\begin{align}
  0&=M_{G^0G^0}^2-\Pi_{G^0G^0}(0)\,, \\
  m_\chi^2&=M_{\chi \chi}^2-\Pi_{\chi \chi}(m_{\chi}^2)\,.
\end{align}
The CP even sector, on the other hand, needs to be
diagonalized. Following ref.~\cite{Brignole:1992uf}, we derive an
eigenstate basis with one-loop correction by diagonalizing
$\bar{M}^2_{ij}$. Using the equations given above, $\bar{M}^2_{ij}$ in
the CP even sector are rewritten as
\begin{align}
  &\bar{M}^2_{hh} =\lambda_{\rm H}v^2+ \Delta_{11}\,, \\
  &\bar{M}^2_{hs} =\lambda_{\rm HS}v v_s+\Delta_{12}\,, \\
  &\bar{M}^2_{ss} =\lambda_{\rm S}v_s^2-\mu_S^{\prime 2}
  +m_{\chi}^2+\Delta_{22}+\Delta \Pi_{\chi\chi}(m_\chi^2)\,,
\end{align}
where
\begin{align}
  &\Delta_{11}\equiv \Pi_{G^0G^0}(0)-\Pi_{hh}(0)\,, \\
  &\Delta_{12}\equiv -\Pi_{hs}(0)\,,
  \label{eq:Delta_12}\\
  &\Delta_{22}\equiv \Pi_{\chi\chi}(0)-\Pi_{ss}(0)\,.
\end{align}
Then the mixing angle for the diagonalization is obtained by
\begin{equation}
O^T \left(
    \begin{array}{cc}
      \bar{M}^2_{hh} & \bar{M}^2_{hh} \\
      \bar{M}^2_{hs} & \bar{M}^2_{ss} 
    \end{array}
    \right)
    O=
    \left(
    \begin{array}{cc}
      \bar{m}_{h_1}^2 & 0 \\
     0  & \bar{m}_{h_2}^2
    \end{array}
    \right)\,,
    \label{eq:diagMCPeven}
\end{equation}
with
\begin{align}
  O=\left(
    \begin{array}{cc}
      \cos \theta & \sin \theta \\
     -\sin \theta  & \cos \theta
    \end{array}
    \right) \,,~~~~~~~
    \tan 2 \theta =-\frac{2\bar{M}^2_{hs}}{\bar{M}^2_{hh}-\bar{M}^2_{ss}}\,.
    \label{eq:OandTan}
\end{align}
The eigenstates $h_1$ $h_2$ are then given by $(h_1,
h_2)^T=O^T(h,s)^T$. We define $h_1$ as the lighter field, which is
identified as the observed Higgs boson.  Their physical masses are
then derived straightforwardly as
\begin{equation}
    m_{h_i}^2=\bar{m}_{h_i}^2-\Delta \Pi_{h_i h_i}(m_{h_i}^2)\,.
\end{equation}

Using the mass eigenstates, the scalar potential can be expanded around the
VEVs. In the mass eigenstate basis, the coefficients for scalar cubic and
quartic couplings, which are relevant for our discussion, are expressed
as
\begin{align}
  {\cal V} \supset
  &\sum_{i} (c_{\chi\chi h_i}\chi^2h_i+c_{G^0G^0h_i}{G^0}^2h_i)
  +\sum_{i\le j \le k}c_{h_i h_j h_k}h_i h_j h_k \nonumber \\
  &+d_{\chi\chi\chi\chi}\chi^4+d_{G^0G^0G^0G^0}{G^0}^4
  +d_{\chi\chi G^0G^0}\chi^2 {G^0}^2 \nonumber \\
  &+\sum_{i \le j}(d_{\chi\chi h_i h_j}\chi^2 h_i h_j 
  +d_{G^0 G^0 h_i h_j}{G^0}^2h_i h_j) 
  +\sum_{i\le j \le k \le m}d_{h_i h_j h_k h_m}h_i h_j h_k h_m \,,
\end{align}
where $i,j,k,m$ are 1 or 2. Additionally, the Yukawa couplings to quarks
are given by
\begin{align}
  {\cal L}_{\rm Yukawa} \supset
  -\sum_i y_{qqh_i} h_i \bar{q}q\,,
\end{align}
where $q$ are quarks.

\section{Scattering Cross Section}
\label{sec:sigma}
\setcounter{equation}{0} 

In this section we compute the SI cross section of $\chi$ DM with
nucleon. To avoid confusing readers, we clarify some terminology
related to our calculation. We will perform the calculation literally
at one-loop level for non-QCD related part.  For QCD part, on the
other hand, the amplitude is derived at the {\it next-to-leading order
  (NLO)} level. Throughout this paper we use the term {\it NLO} or
{\it leading order (LO)} in terms of order of QCD strong coupling
$\alpha_s$. For example, {\it LO} contains one-loop diagrams for
$\chi$--$g$ scattering. However, the gluon contributions to the
effective scalar coupling are ${\cal
  O}(\alpha_s^0)$~\cite{Hisano:2015rsa,Hisano:2010fy,Hisano:2010ct}. That
is why we use the term {\it LO} for such $\chi$--$g$ processes, and
similar discussion is applied for {\it NLO}. As we will see, the gluon
contributions become important in some parameter space.

\subsection{Formalism}

We briefly summarize the formalism for the calculation of the SI
scattering cross section of a real scalar DM with nucleon based on
refs.~\cite{Hisano:2015rsa,Hisano:2015bma}. Using the formalism in
ref.~\cite{Hisano:2015rsa}, we calculate the scattering amplitude at
the NLO in QCD.

The effective Lagrangian relevant for the scattering process is
\begin{align}
  {\cal L}_{\rm eff}
&=
\sum_{i=q,G}C^i_{\rm S} {\cal O}^i_{\rm S}
+\sum_{i=q,G}
C^i_{\rm T} {\cal O}^i_{\rm T}\,,
\end{align}
where $C^i_{\rm S}$ and $C^i_{\rm T}$ are the Wilson coefficients and
the operators ${\cal O}^i_{\rm S}$ and ${\cal O}^i_{\rm T}$ are given
by
\begin{align}
  {\cal O}^q_{\rm S}&\equiv m_q \chi^2 \bar{q}q\,,\nonumber \\
 {\cal O}^G_{\rm S}&\equiv \frac{\alpha_s}{\pi}
\chi^2 G^a_{\mu\nu}G^{a\mu\nu}\,,
\nonumber \\
 {\cal O}^i_{\rm T}&\equiv \frac{1}{m_\chi^2}
\chi i\partial^\mu i\partial^\nu\chi {\cal O}^i_{\mu\nu}\,.
\label{eq:}
\end{align}
Here $G^a_{\mu\nu}$ represents the field strength tensor of gluon
field and the quark masses are denoted as $m_q$.  The operators ${\cal
  O}^q_{\mu\nu}$ and ${\cal O}^G_{\mu\nu}$ are the twist-2 operators
of quarks and gluon, respectively, which are defined by
\begin{align}
 {\cal O}^q_{\mu\nu}&\equiv \frac{1}{2}\overline{q}i\biggl(
D_\mu^{}\gamma_\nu^{} +D_\nu^{}\gamma_\mu^{}-\frac{1}{2}g_{\mu\nu}^{}
\Slash{D}\biggr)q\,,\nonumber \\
{\cal O}^G_{\mu\nu}&\equiv 
G^{a\rho}_{\mu} G^{a}_{\nu\rho}-\frac{1}{4}g_{\mu\nu}^{}
G^a_{\rho\sigma}G^{a\rho\sigma}\,,
\label{eq:twist2def}
\end{align}
with $D_\mu$ the covariant derivative.  Then the SI scattering cross section
of $\chi$ with nucleon $N$ is obtained as~\cite{Hisano:2015rsa,Hisano:2015bma}
\begin{align}
  \sigma^N_{\text{SI}} =\frac{1}{\pi}\biggl(\frac{m_N}{m_\chi+m_N}\biggr)^2
\left|f^N_{\text{scalar}}+f^N_{\text{twist2}}\right|^2\,,
\end{align}
where $m_N$ is the nucleon mass, and $f^N_{\rm scalar}$ and $f^N_{\rm
twist2}$ are given by
\begin{align}
&\frac{f^N_{\rm scalar}}{m_N}=
  \sum_{q=u,d,s}C^q_{\text{S}}(\mu_{\text{had}}) f_{Tq}^N
-\frac{8}{9}C^G_{\text{S}}(\mu_{\text{had}})f_{Tg}^N\,,
\\
&\frac{f^N_{\rm twist2}}{m_N}=
\frac{3}{4}\sum_{q}\Bigl[C^q_{\text{T}}(m_Z)
[q^{N}(2;m_Z)+\bar{q}^{N}(2;m_Z)]
-C^G_{\text{T}}(m_Z)g^{N}(2; m_Z)\Bigr]\,.
\end{align}
Here $f_{Tq}^N$, $f_{Tg}^N$, $q^{N}(2;m_Z)$, $\bar{q}^{N}(2;m_Z)$, and
$g^{N}(2;m_Z)$ are the matrix elements of the effective operators in
nucleon state. $\mu_{\rm had}$ is the hadronic scale ({\it i.e.},
around 1 GeV), and $m_Z$ is the $Z$ boson mass. The numerical values for
these quantities are given in ref.~\cite{Hisano:2015rsa}\footnote{To be strict,
  $f_{Tq}^N$ corresponds to $f_{Tq}^{(N)}$ in ref.~\cite{Hisano:2015rsa}
  and so on. We have additionally introduced $f_{Tg}^N$ defined as
  $(-8/9)f_{Tg}^N\equiv \langle N|\frac{\alpha_s}{\pi}
  G^a_{\mu\nu}G^{a\mu\nu}|N\rangle /m_N$ evaluated at three flavors,
  which leads to $f_{Tg}^N=1-\sum_{q=u,d,s} f_{Tq}^N+{\cal
    O}(\alpha_s)$.}  based on the QCD lattice
    simulation~\cite{Young:2009zb, Oksuzian:2012rzb} and CETEQ-Jefferson
    Lab collaboration~\cite{Owens:2012bv}. As we will see, the contribution to
the twist-2 type operators is negligibly small. Therefore, the SI
cross section is  determined by the scalar-type interactions.

\subsection{Wilson coefficients}

\begin{figure}[t]
 \begin{center}
     \includegraphics[scale=0.7]{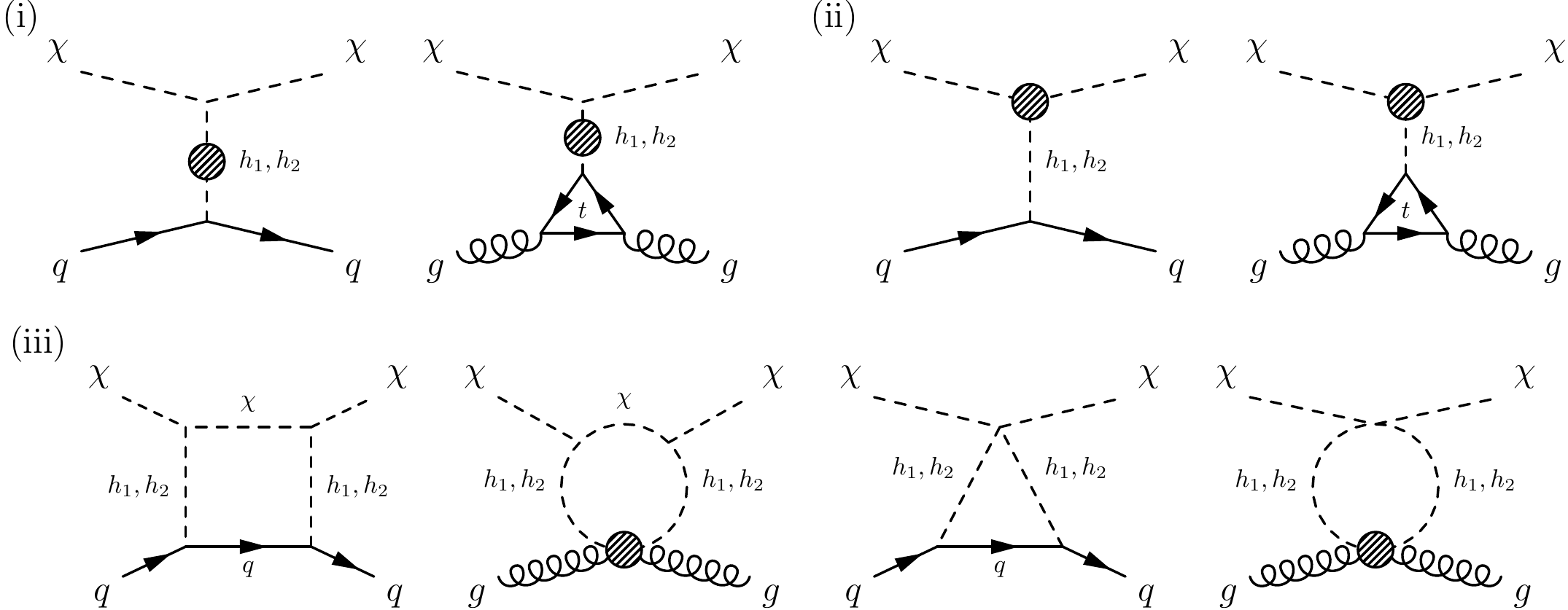}
  \caption{Feynman diagrams for $\chi$--$q$ and $\chi$-$g$ scattering
    processes.}
  \label{fig:scat}
 \end{center}
\end{figure}

First of all, we derive the effective Lagrangian from full theory by
matching at the weak scale denoted as $\mu_{W}\simeq m_Z$. As
described in Introduction, the tree-level amplitudes for $\chi q \to
\chi q$ are cancelled in the non-relativistic limit. Therefore
loop-level calculations are necessary to evaluate the scattering
amplitude in the limit. There are three types of diagrams shown in
Fig.\,\ref{fig:scat}; (i) Self-energies, (ii) Vertex corrections, and
(iii) Box and triangle diagrams.\footnote{ The NLO diagrams for QCD
  part are not depicted for simplicity, but these are taken into
  account in the numerical study.}

The most part of the computation for the diagrams (i) has already been
done in the previous section. Let us apply the results to the
scattering process. The diagrams (i) give rise to scalar-type
interactions. The matching at the weak scale gives
\begin{align}
  &C_{\rm S}^q(\mu_W)\Bigr|_{\rm self}=-\frac{1}{2m^2_{h_1}m^2_{h_2}vv_s}
  \Bigl[
    \Delta_{12}(\sin^2\theta \, m_{h_1}^2+\cos^2\theta \, m_{h_2}^2)
    \nonumber \\
    &~~~~~~~~~~~~~~~~~~~~~~~~~~~~~~~~
    +\tilde{\Delta}_{22}\sin \theta \cos \theta (m_{h_1}^2-m_{h_2}^2)
    \Bigr]\,, \\
  &C_{\rm S}^G(\mu_W)\Bigr|_{\rm self}=-\frac{1}{12}
  \left[1+\frac{11\alpha_s}{4\pi}\right]C_{\rm S}^q(\mu_W)\Bigr|_{\rm self}\,,
  \label{eq:CG_i}
\end{align}
for $q=u,d,s,c,b$ where $\Delta_{12}$ is given in
Eq.\,\eqref{eq:Delta_12}, and $\tilde{\Delta}_{22}$ is given by
\begin{equation}
    \tilde{\Delta}_{22}=\frac{T_s}{v_s}-\Pi_{ss}(0)\,.
\end{equation}

The computation of the diagrams (ii) is rather simple. These diagrams
give vertex corrections to the $\chi$-$\chi$-$h_i$ couplings. Denoting
them as $\Delta c_{\chi\chi h_i}$ (collected in Appendix), the Wilson
coefficients are obtained as
\begin{align}
  &C_{\rm S}^q(\mu_W)\Bigr|_{\rm vert}=  \frac{1}{m_{h_1}^2m_{h_2}^2v}
  \Bigl[
    \Delta c_{\chi\chi h_1}\cos\theta\, m_{h_2}^2 +
    \Delta c_{\chi\chi h_2}\sin\theta\, m_{h_1}^2
    \Bigr]\,, \\
   &C_{\rm S}^G(\mu_W)\Bigr|_{\rm vert}=-\frac{1}{12}
  \left[1+\frac{11\alpha_s}{4\pi}\right]C_{\rm S}^q(\mu_W)\Bigr|_{\rm vert}\,,
\end{align}
for $q=u,d,s,c,b$.  It is noted that, although both $C^q_{\rm
  S}(\mu_W)|_{\rm self}$ and $C^q_{\rm S}(\mu_W)|_{\rm vert}$ have the
renormalization scale dependence, the dependence
is cancelled in $C^q_{\rm S}(\mu_W)|_{\rm self}+C^q_{\rm
  S}(\mu_W)|_{\rm vert}$ as expected.

The $\chi$--$q$ scattering process drawn in the diagrams (iii) gives
both scalar and twist-2 type contributions. However, the resultant
Wilson coefficients are proportional to $y_{qqh_i}^2 \propto m_q^2$
($q=u,d,s,c,b$), thus they are negligibly small. For the $\chi$--$g$
scattering, on the other hand, the top loop diagram should be taken
into account since there is no such suppression. They can be
calculated easily in Fock-Schwinger gauge (see Appendix for details) 
and the resultant expressions are
\begin{align}
  C_{\rm S}^G(\mu_W)\Bigr|_{\rm box+tri}=\frac{1}{m_{\chi}^4}
  \sum_{i\le j}y_{tth_i}y_{tth_j}
  \left[c_{\chi\chi h_i}c_{\chi\chi h_j}J_{\rm box}^{ij}
  +m_\chi^2d_{\chi\chi h_i h_j}J_{\rm tri}^{ij}\right]\,.
\end{align}
We refer to the terms proportional to $J_{\rm box}^{ij}$ and $J_{\rm
  tri}^{ij}$ as `box' type and `triangle' type, respectively.  Note
that there is no $m_\chi$ dependence in $J_{\rm tri}^{ij}/m_\chi^2$,
which is obvious from the corresponding diagrams. It is found
numerically that the contribution to the amplitude from the box-type
diagrams is much smaller than that from the triangle-type diagrams in
the parameter space we are interested in. Here we have ignored NLO
contributions and we will treat it as a theoretical uncertainty as in
ref.~\cite{Hisano:2015rsa}. This is because it is expected to be
suppressed compared to the other NLO contributions.

To summarize, the weak scale matching gives
\begin{align}
  C_{\rm S}^q(\mu_W)&=
  C_{\rm S}^q(\mu_W)\Bigr|_{\rm self}+
  C_{\rm S}^q(\mu_W)\Bigr|_{\rm vert}\,, \\
  C_{\rm S}^G(\mu_W)&=
  C_{\rm S}^G(\mu_W)\Bigr|_{\rm self}+
  C_{\rm S}^G(\mu_W)\Bigr|_{\rm vert}+
  C_{\rm S}^G(\mu_W)\Bigr|_{\rm box+tri}\,.
\end{align}
The Wilson coefficients at the hadronic scale $\mu_{\rm had}$ are
obtained by the renormalization group equations, along with the
matching at bottom and charm mass scales, consistently at the NLO in
QCD~\cite{Hisano:2015rsa}.

\subsection{Numerical results}

\begin{figure}[t]
 \begin{center}
   \includegraphics[scale=0.75]{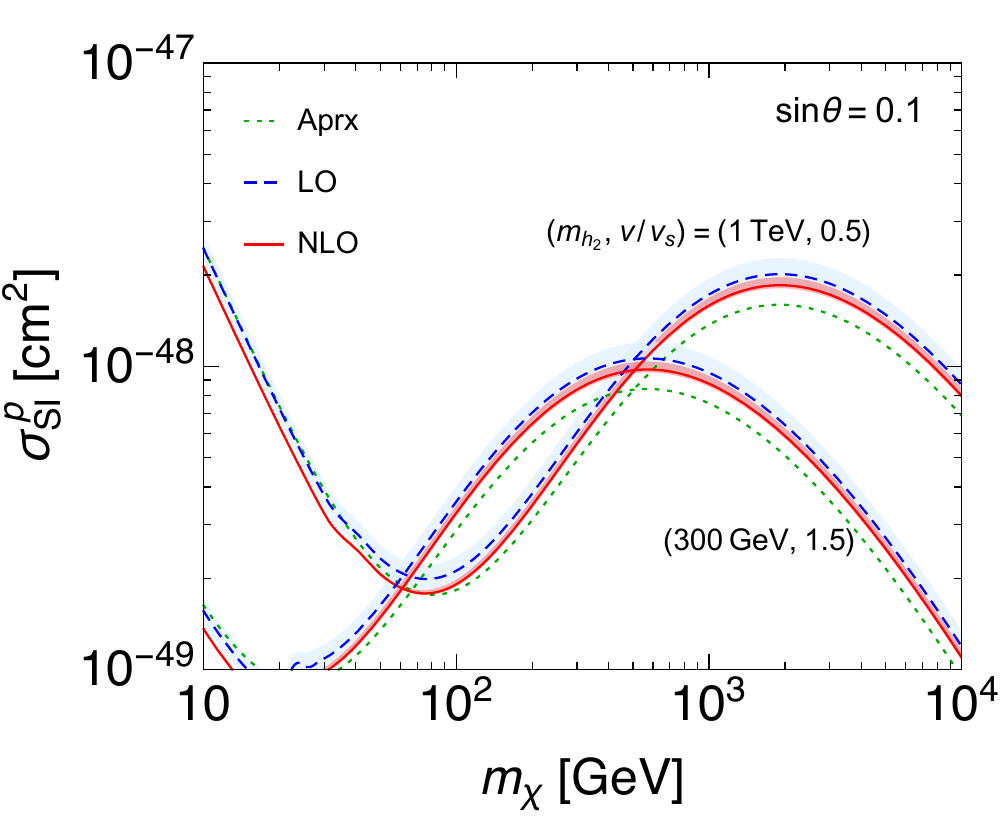}~~
   \includegraphics[scale=0.75]{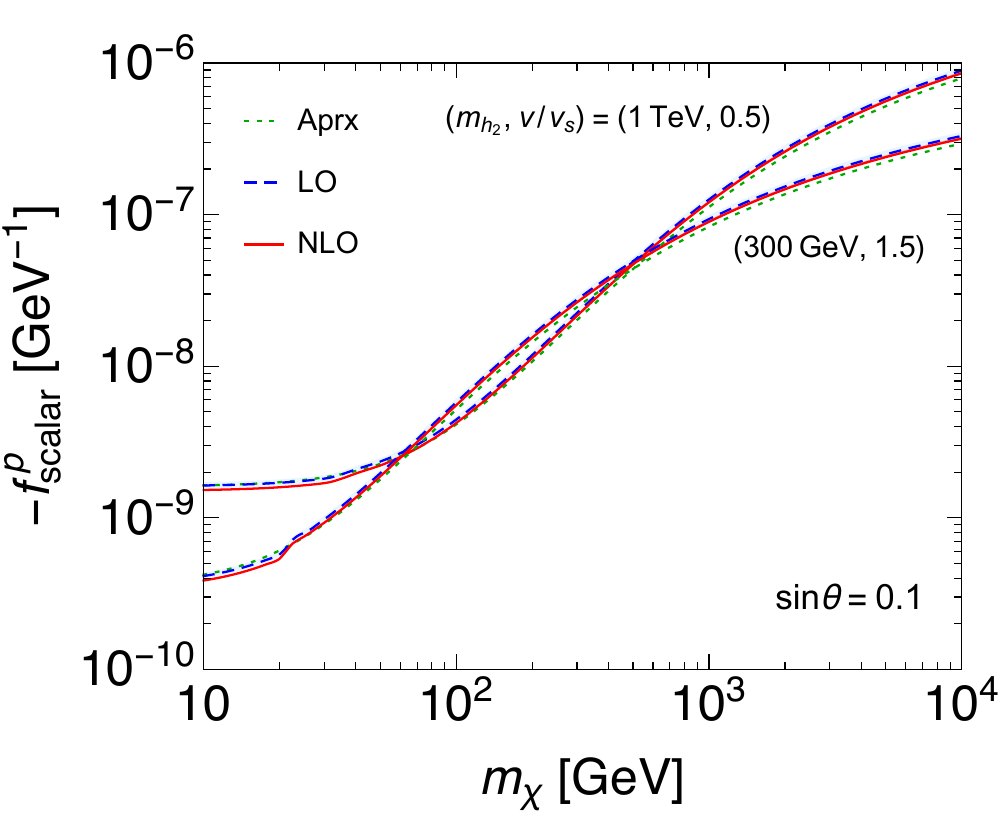}
  \caption{(Left) Spin-independent cross section of DM with proton for
    two sample parameter points, $(m_{h_2},v/v_s)=(300~{\rm GeV},1.5)$
    and $(1~{\rm TeV},0.5)$, for $\sin \theta=0.1$.  The numerical
    evaluations with the approximate expression
    Eq.\,\eqref{eq:apprxf^N}, LO and NLO in QCD are shown as `Aprx'
    (green dotted), `LO' (blue dashed), and `NLO' (red solid). For LO
    and NLO results, the perturbative errors are also shown in shaded
    region. (Right) (Minus) scalar-type effective coupling $f^p_{\rm
      scalar}$ for the same sample parameter points. The line legends
    are the same as left figure.}
  \label{fig:sigma}
 \end{center}
\end{figure}

Now we are ready to show the numerical results. The SI cross section
of $\chi$ DM with proton is plotted in Fig.\,\ref{fig:sigma}. We have
computed the SI cross section at the NLO in QCD. For comparison, the
results at the LO in QCD and the results obtained by an approximate
expression for $f^N_{\rm scalar}$ given by
\begin{align}
  f^N_{\rm scalar}/m_N \approx
  \left[\frac{2}{9}+\frac{7}{9}\sum_{q=u,d,s}f_{Tq}^N
    \right]C_{\rm S}^q(\mu_W)
  -\frac{8}{9}f_{Tg}^N C_{\rm S}^G(\mu_W)\Bigr|_{\rm box+tri}
  \,,
  \label{eq:apprxf^N}
\end{align}
are shown as well. Here $f_{Tg}^N$ is given at the LO. The deviations
of the LO and the approximate results from the NLO results are both
${\cal O}(10\%)$.\footnote{For example, the deviation of the LO and
  the approximation is 9\% and $ -14$\%, respectively, for
  $(m_{h_2},v/v_s)=(1~{\rm TeV},0.5)$, $\sin \theta=0.1$ and
  $m_{\chi}=1~{\rm TeV}$. Such behavior is observed in a similar DM
  model studied in ref.~\cite{Endo:2015nba}.}  Perturbative errors at
the LO results are ${\cal O}(10\%)$, which are reduced to a few \% at
the NLO level as expected. We have also checked that the errors due to
the input parameters are ${\cal O}(10\%)$, and that the uncertainty
due to ignoring the NLO contribution in the diagrams (iii) is less
than 1\%. The behavior of the cross section is understood from the
scalar-type effective coupling $f^N_{\rm scalar}$, which is plotted in
the right of Fig.\,\ref{fig:sigma}.  It is found that in large
$m_\chi$ region the contributions from the diagrams (iii) are
subdominant in $f^N_{\rm scalar}$, consequently, the effective
coupling is determined by the diagrams (i) and (ii), {\it i.e.}, the
self-energies and vertex corrections that have a logarithmic
dependence on $m_\chi$ for large $m_\chi$ (see
Eq.\,\eqref{eq:self+vert|largemchi} in Appendix). Thus $|f^N_{\rm
  scalar}|$ increases as $\log m_\chi+$const. In small $m_\chi$
region, on the contrary, the diagrams (iii) dominate over the diagrams
(i) and (ii) which are suppressed by $m_\chi^2$ (see
Eq.\,\eqref{eq:self+vert|smallmchi} in Appendix). This is why the
cross section is not suppressed in small DM mass region. This
turn-over in the effective coupling happens roughly around
$m_{\chi}\sim {\cal O}(10\,\mathchar`-\,10^3\,{\rm GeV})$.

\begin{figure}[t]
  \begin{center}
   \includegraphics[scale=0.62]{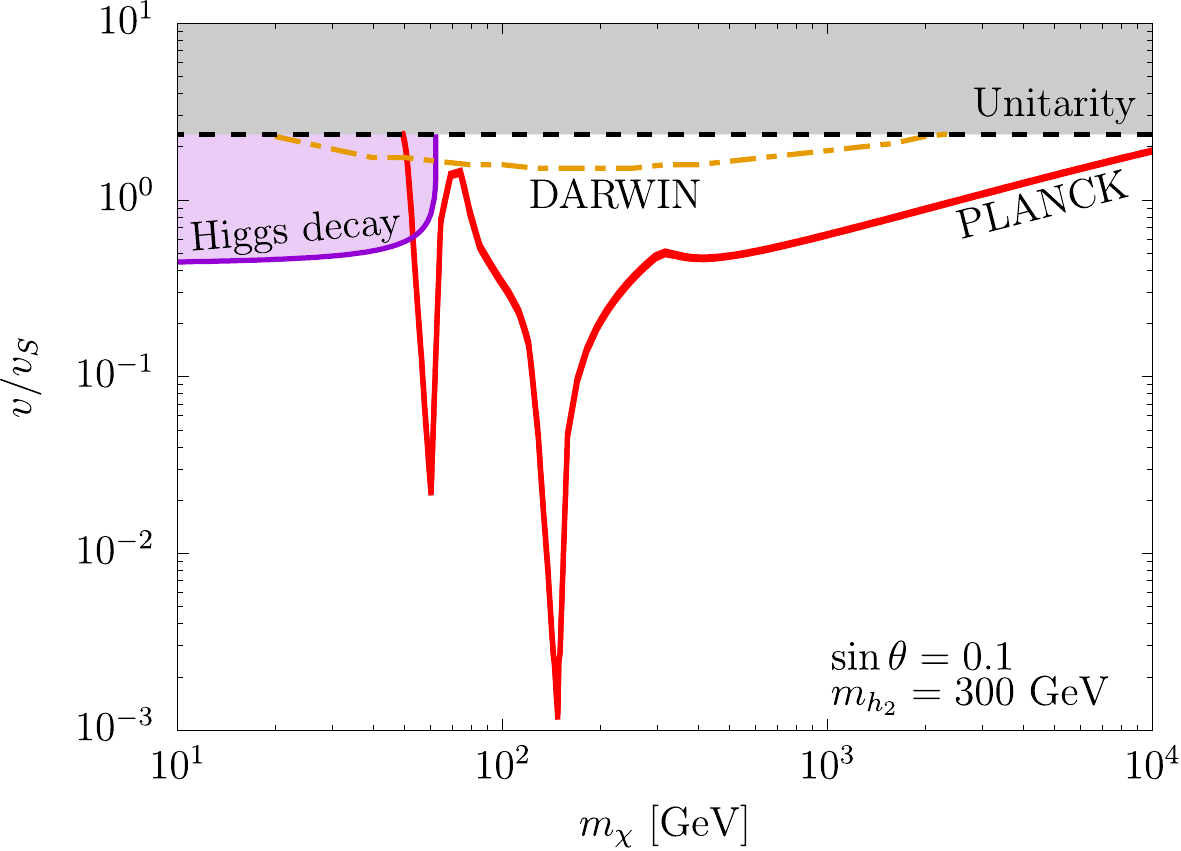}
   \includegraphics[scale=0.62]{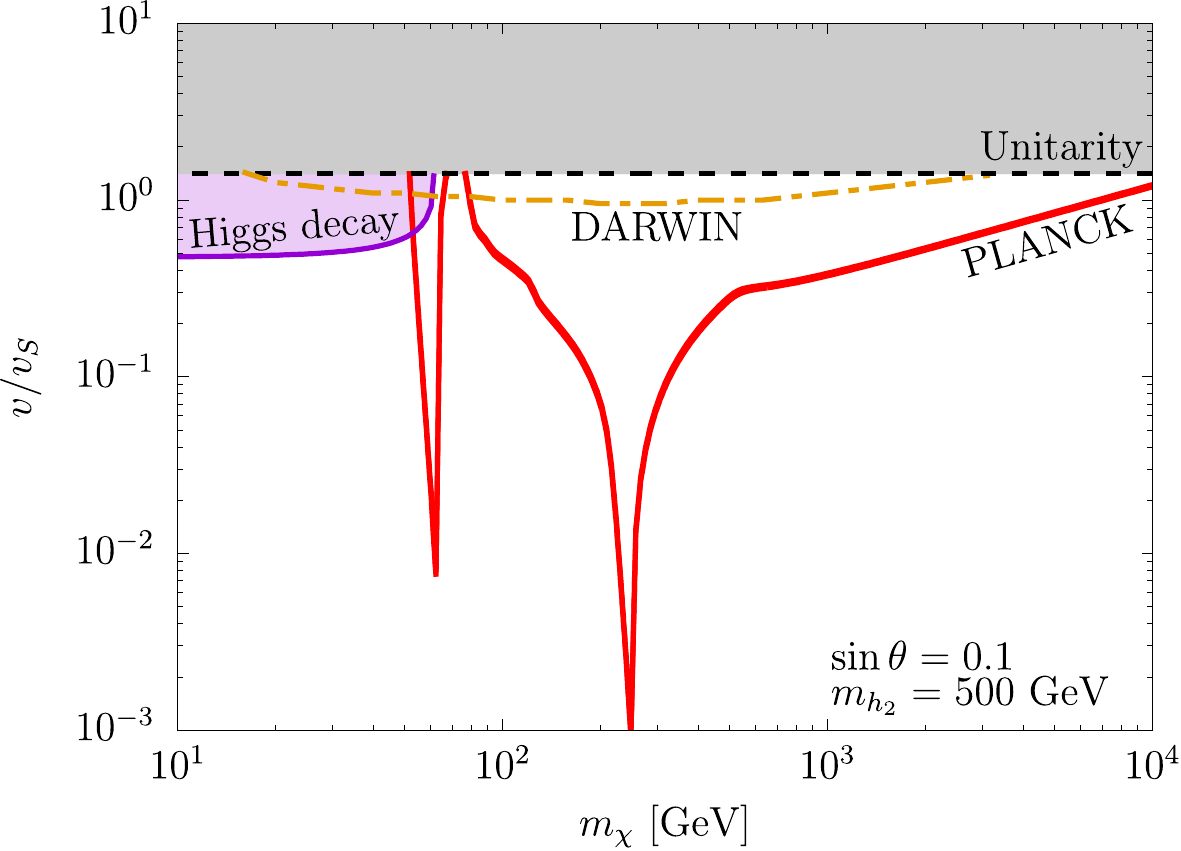}
   \includegraphics[scale=0.62]{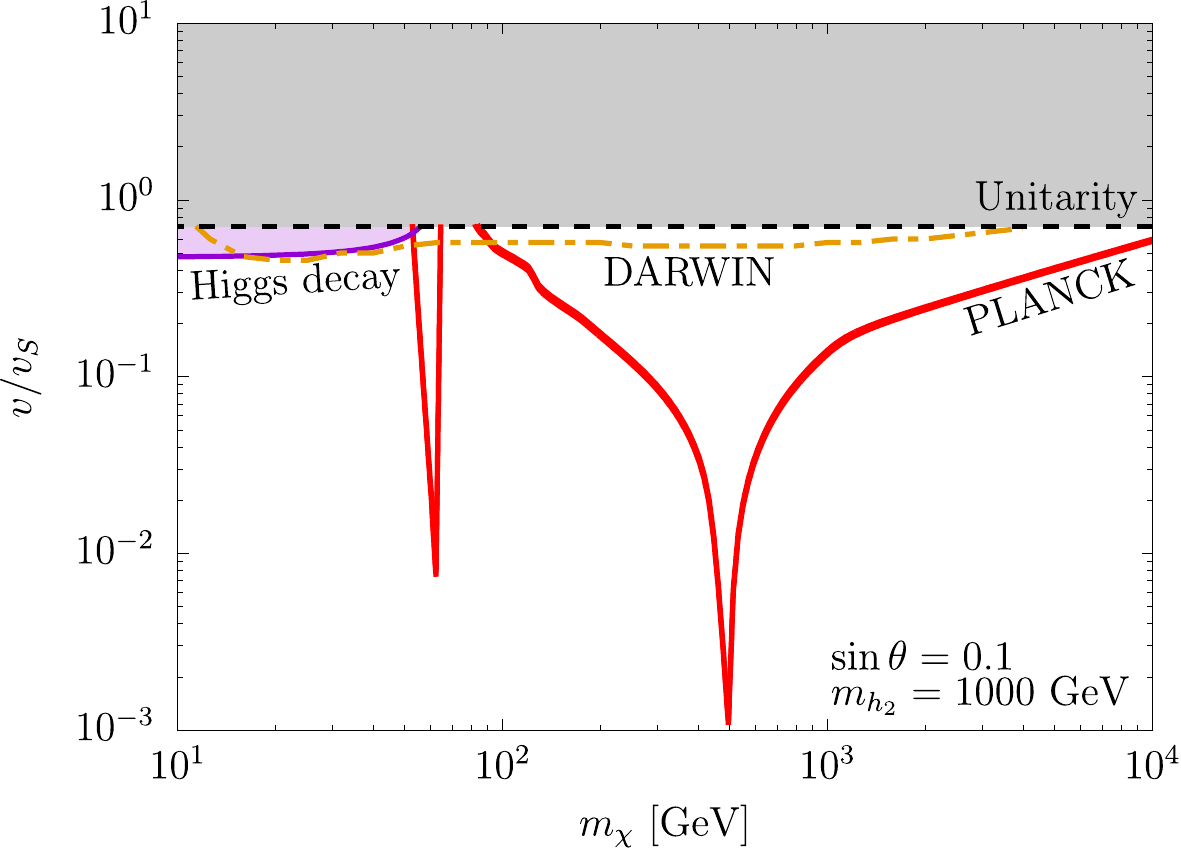}
   \caption{ Exclusion limits on the model from the Higgs decay
     (purple) and the perturbative unitarity (grey). The red band
     represents the parameter space which can reproduce the observed
     relic abundance in the thermal freeze-out scenario. The future
     prospect of the DARWIN experiment is also shown as the orange
     dot-dashed line~\cite{Aalbers:2016jon} (region above the line can
     be probed).}
  \label{fig:limit}
 \end{center}
\end{figure}

Fig.\,\ref{fig:limit} shows various constraints on the model.  The
purple region is excluded by the constraint from the SM-like Higgs
decay where the Higgs signal strength at the LHC
$\mu=1.09^{+0.11}_{-0.10}$ has been translated into the constraint on
the Higgs invisible decay as
$\mathrm{Br}(h_1\to\mathrm{inv})\leq0.11$~\cite{Khachatryan:2016vau}.
The grey region is excluded by the perturbative unitarity bound
$\lambda_{\rm S}\leq 8\pi/3$~\cite{Chen:2014ask}.  The red band
represents the parameter space which can reproduce the observed DM
relic abundance within $3\sigma$ range of the PLANCK Collaboration
data~\cite{Ade:2015xua} in the thermal freeze-out scenario.  It has
been found that since the SI cross section is so suppressed in a wide
range of parameter space that there is no substantial constraint on
the model from the current experimental limit of the XENON1T
experiment~\cite{Aprile:2018dbl}. To be concrete, the region excluded
by XENON1T is always in the unitarity bound.  In the plots, the future
reach of the DARWIN experiment~\cite{Aalbers:2016jon} is also shown in
the orange dot-dashed line, assuming that all the DM abundance is
composed by $\chi$.  It indicates that a part of the parameter space
($70~\mathrm{GeV}\lesssim m_\chi\lesssim 100~\mathrm{GeV}$) can be
probed by the DARWIN experiment, where the thermal relic DM scenario
can reproduce all the DM abundance.  We have seen a similar indication
for larger $\sin \theta$ values. Therefore the future direct detection
experiments will be able to probe a part of the thermal freeze-out
scenario with $m_{\chi}\sim 100~{\rm GeV}$.

\section{Conclusion}

We have studied the detectability of DM in a model where a complex
singlet scalar is added to the SM. In the model, a softly broken
global $U(1)$ symmetry has been assumed, and a would-be
Nambu-Goldstone boson $\chi$ becomes a candidate for DM due to a
remnant $\mathbb{Z}_2$ symmetry. Since $\chi$ interacts with the SM
particles via so-called Higgs portal coupling, it would be possible to
directly detect $\chi$ DM via $\chi$--nucleon scattering. It is known,
however, that the tree-level scattering amplitude vanishes in
non-relativistic limit. Thus we have taken into account the scattering
process at one-loop level for non-QCD part. For QCD effect, on the
other hand, the scattering amplitude has been calculated at the
next-to-leading order in QCD strong coupling systematically to reduce
the theoretical uncertainties. It has been found that the predicted SI
cross section is small in a wide range of the parameter space of the
model, and there is no substantial bound from the current direct
detection experiments. However, a part of the parameter space, which
includes canonical thermal relic scenario accounting for the present
DM abundance, will be probed in the future direct detection experiment
DARWIN.

\vspace{0.5cm}

\noindent {\it Acknowledgements}:
This work was supported by JSPS KAKENHI Grant Numbers JP17H05402,
JP17K14278, JP17H02875 and JP18H05542 (K.I.).
T.T. acknowledges support from JSPS Fellowships for Research Abroad.

\vspace{0.5cm}

\noindent {\it Note added}: While completing our paper, we found that
ref.~\cite{Azevedo:2018exj} studied the direct detection of dark
matter in the same model. We agree qualitatively with their results in
large dark matter mass region, as well as in the other dark matter
mass region if the diagrams (iii) are ignored. On the other hand, we
have seen a different behavior in low dark matter mass region. This is
due to the diagrams (iii), which are discarded in their
study. Although these diagrams are irrelevant for the estimation of
the current bound from the XENON1T experiment, it will be important
for the future study of this model in the direct detection
experiments.

\appendix

\section{Loop functions}

The loop functions are basically expressed by so-called $A_0$ function
and $B_0$ function (and their derivatives),
\begin{align}
 iA_0(m^2)&=\int \frac{d^Dq}{(2\pi)^D} \frac{1}{q^2-m^2}\,, \\ 
  iB_0(p^2;m_1^2,m_2^2)&=
  \int \frac{d^Dq}{(2\pi)^D} \frac{1}{(q^2-m_1^2)((q+p)^2-m_2^2)}\,.\\
  iC^{ij}&=\int \frac{d^Dq}{(2\pi)^D}
  \frac{1}{((q+P)^2-m_\chi^2)(q^2-m_{h_i}^2)(q^2-m_{h_j}^2)}\,, \\
  iC^{\chi i}&=\int \frac{d^Dq}{(2\pi)^D}
  \frac{1}{((q+P)^2-m_{h_i}^2)(q^2-m_{\chi}^2)(q^2-m_{\chi}^2)}\,,
\end{align}
where $P^2=m_{\chi}^2$ in the last two functions.  Here the divergent
pieces are subtracted in the $\overline{\rm MS}$ renormalization
scheme implicitly. We use the analytic expressions for the loop
functions, which are numerically checked by using
LoopTools~\cite{Hahn:1998yk}.

\subsection{Tadpoles}

The tadpoles $T_{h_i}$ are given by $A_0$ function as,
\begin{align}
  &T_{h_1}=
  3 c_{h_1h_1h_1} A_0(m_{h_1}^2)+c_{h_1 h_2 h_2}A_0(m_{h_2}^2)
  +c_{\chi\chi h_1}A_0(m_{\chi}^2)-y_{tth_1}m_t A_0(m_t^2)\,,\\
  &T_{h_2}=
  3 c_{h_2h_2h_2} A_0(m_{h_2}^2)+c_{h_1 h_1 h_2}A_0(m_{h_1}^2)
  +c_{\chi\chi h_2}A_0(m_{\chi}^2)-y_{tth_2}m_t A_0(m_t^2)\,. 
\end{align}
Then $T_h$ and $T_s$ are obtained by rotating with the orthogonal matrix
$O$ given in Eq.~(\ref{eq:OandTan}) as $(T_h, T_s)^T=O (T_{h_1},T_{h_2})^T$.

\subsection{Self-energies}

The one-loop contributions to the self-energies $\Pi_{h_1h_1}$,
$\Pi_{h_2h_2}$, and $\Pi_{h_1h_2}$ are listed below: 
\begin{align}
  &\Pi_{h_1h_1}(p^2)=
  2 c_{\chi \chi h_1}^2 B_0(p^2;m_\chi^2,m_\chi^2)
  +6 c_{G^0G^0h_1}^2B_0(p^2;0,0) \nonumber \\
  &+18c_{h_1h_1h_1}^2B_0(p^2;m_{h_1}^2,m_{h_1}^2)
  +2c_{h_1h_2h_2}^2B_0(p^2;m_{h_2}^2,m_{h_2}^2)
  +4c_{h_1h_1h_2}^2B_0(p^2;m_{h_1}^2,m_{h_2}^2) \nonumber \\
  &+2 d_{\chi\chi h_1 h_1} A_0(m_{\chi}^2)
  +12 d_{h_1 h_1 h_1 h_1} A_0(m_{h_1}^2)
  +2 d_{h_1 h_1 h_2 h_2} A_0(m_{h_2}^2) \nonumber \\
  &-4y_{tth_1}^2(A_0(m_t^2)+(2m_t^2-p^2/2)B_0(p^2;m_t^2,m_t^2))\,, \\
  &\Pi_{h_2h_2}(p^2)=
  2 c_{\chi \chi h_2}^2 B_0(p^2;m_\chi^2,m_\chi^2)
  +6 c_{G^0G^0h_2}^2B_0(p^2;0,0) \nonumber \\
  &+18c_{h_2h_2h_2}^2B_0(p^2;m_{h_2}^2,m_{h_2}^2)
  +2c_{h_1h_1h_2}^2B_0(p^2;m_{h_1}^2,m_{h_1}^2)
  +4c_{h_1h_2h_2}^2B_0(p^2;m_{h_1}^2,m_{h_2}^2) \nonumber \\
  &+2 d_{\chi\chi h_2 h_2} A_0(m_{\chi}^2)
  +12 d_{h_2 h_2 h_2 h_2} A_0(m_{h_2}^2)
  +2 d_{h_1 h_1 h_2 h_2} A_0(m_{h_1}^2) \nonumber \\
  &-4y_{tth_2}^2(A_0(m_t^2)+(2m_t^2-p^2/2)B_0(p^2;m_t^2,m_t^2))\,, \\
  &\Pi_{h_1h_2}(p^2)=
  2 c_{\chi \chi h_1}c_{\chi \chi h_2} B_0(p^2;m_\chi^2,m_\chi^2)
  +6 c_{G^0G^0h_1}c_{G^0G^0h_2}B_0(p^2;0,0) \nonumber \\
  &+6c_{h_1h_1h_1}c_{h_1h_1h_2}B_0(p^2;m_{h_1}^2,m_{h_1}^2)
  +6c_{h_2h_2h_2}c_{h_1h_2h_2}B_0(p^2;m_{h_2}^2,m_{h_2}^2) \nonumber \\
  &+4c_{h_1h_1h_2}c_{h_1h_2h_2}B_0(p^2;m_{h_1}^2,m_{h_2}^2) \nonumber \\
  &+d_{\chi\chi h_1 h_2} A_0(m_{\chi}^2)
  +3d_{h_1 h_1 h_1 h_2} A_0(m_{h_1}^2)
  +3d_{h_1 h_2 h_2 h_2} A_0(m_{h_2}^2)\nonumber \\
  &-4y_{tth_1}y_{tth_2}(A_0(m_t^2)+(2m_t^2-p^2/2)B_0(p^2;m_t^2,m_t^2))\,.
\end{align}
Then, $\Pi_{hh}$, $\Pi_{ss}$, and $\Pi_{hs}$ are obtained similarly to
Eq.\,\eqref{eq:diagMCPeven}.

\subsection{Vertex corrections}

The corrections to the cubic couplings $c_{\chi \chi h_i}$ are given by
\begin{align}
  \Delta c_{\chi\chi h_i}=
  \Delta c_{\chi\chi h_i}^{\rm c}+\Delta c_{\chi\chi h_i}^{\rm t}\,,
\end{align}
with
\begin{align}
  - \Delta c_{\chi\chi h_1}^{\rm c} =&
  6 d_{\chi\chi h_1 h_1}c_{h_1h_1h_1}B_0(0;m_{h_1}^2,m_{h_1}^2)+
  2 d_{\chi\chi h_1 h_2}c_{h_1h_1h_2}B_0(0;m_{h_1}^2,m_{h_2}^2)\nonumber \\
  &+2 d_{\chi\chi h_2 h_2}c_{h_1h_2h_2}B_0(0;m_{h_2}^2,m_{h_2}^2)+ 
  12d_{\chi\chi\chi\chi}c_{\chi\chi h_1}B_0(0;m_{\chi}^2,m_{\chi}^2)\nonumber \\
  &+8d_{\chi\chi h_1 h_1}c_{\chi\chi h_1}B_0(0;m_{\chi}^2,m_{h_1}^2)+ 
  4d_{\chi\chi h_1 h_2}c_{\chi\chi h_2}B_0(0;m_{\chi}^2,m_{h_2}^2)\,,
  \\
  - \Delta c_{\chi\chi h_2}^{\rm c} =&
  6 d_{\chi\chi h_2 h_2}c_{h_2h_2h_2}B_0(0;m_{h_2}^2,m_{h_2}^2)+
  2 d_{\chi\chi h_1 h_2}c_{h_1h_2h_2}B_0(0;m_{h_1}^2,m_{h_2}^2)\nonumber \\
  &+2 d_{\chi\chi h_1 h_1}c_{h_1h_1h_2}B_0(0;m_{h_2}^2,m_{h_2}^2)+ 
  12d_{\chi\chi\chi\chi}c_{\chi\chi h_2}B_0(0;m_{\chi}^2,m_{\chi}^2)\nonumber \\
  &+8d_{\chi\chi h_2 h_2}c_{\chi\chi h_2}B_0(0;m_{\chi}^2,m_{h_2}^2)+
  4d_{\chi\chi h_1 h_2}c_{\chi\chi h_1}B_0(0;m_{\chi}^2,m_{h_1}^2)\,, \\
  - \Delta c_{\chi\chi h_1}^{\rm t} =&
  12c_{\chi\chi h_1}^2 c_{h_1h_1h_1}C^{11}
  +8c_{\chi\chi h_1}c_{\chi\chi h_2} c_{h_1h_1h_2}C^{12}
  +4c_{\chi\chi h_2}^2 c_{h_1h_2h_2}C^{22}\nonumber \\
  &+4c_{\chi\chi h_1}^3C^{\chi 1}
  +4c_{\chi\chi h_1}c_{\chi\chi h_2}^2C^{\chi 2}\,,
   \\
  - \Delta c_{\chi\chi h_2}^{\rm t} =&
  12c_{\chi\chi h_2}^2 c_{h_2h_2h_2}C^{22}
  +8c_{\chi\chi h_1}c_{\chi\chi h_2} c_{h_1h_2h_2}C^{12}
  +4c_{\chi\chi h_1}^2 c_{h_1h_1h_2}C^{11}\nonumber \\
  &+4c_{\chi\chi h_2}^3C^{\chi 2}
  +4c_{\chi\chi h_1}^2c_{\chi\chi h_2}C^{\chi 1}\,.
\end{align}
As described in the main text of the paper, the renormalization scale
dependence is cancelled in the Wilson coefficients, $C^q_{\rm
  S}(\mu_W)|_{\rm self}+C^q_{\rm S}(\mu_W)|_{\rm vert}$.

It would be helpful to see how it behaves in small and large $m_\chi$
limit. For $m_\chi \to 0$ limit, we have found that it is proportional
to $m_\chi^2$. For example,\footnote{We omit the expression for
  non-zero $m_{h_1}$ and $m_{h_2}$ since it is too lengthy. (It is
  similar for large $m_\chi$ case.)}
\begin{align}
  &C^q_{\rm S}(\mu_W)|_{\rm self}+C^q_{\rm S}(\mu_W)|_{\rm vert} \nonumber \\
  &\to
  \left\{
  \begin{array}{ll}
    \frac{2m_\chi^2m_{h_2}^4}{v v_s^2}
    \sin\theta \cos^4\theta (v \cos\theta-v_s \sin\theta )
    & ({\rm for}~m_{h_1}=0)\\
    - \frac{2m_\chi^2m_{h_1}^4}{v v_s^2}
    \sin^4\theta \cos\theta (v \sin\theta+v_s \cos\theta )
    & ({\rm for}~m_{h_2}=0)
  \end{array}\right. \,.
  \label{eq:self+vert|smallmchi}
\end{align}
In large $m_\chi$ limit, on the other hand, its absolute value increases
logarithmically, {\it e.g.},
\begin{align}
  &C^q_{\rm S}(\mu_W)|_{\rm self}+C^q_{\rm S}(\mu_W)|_{\rm vert} \nonumber \\
  &\to
  \left\{
  \begin{array}{ll}
    \frac{2\left[\log (m_\chi^2/m_{h_2}^2)+2\right]m_{h_2}^6}{v v_s^2}
    \sin\theta \cos^4\theta (v \cos\theta-v_s \sin\theta )
    & ({\rm for}~m_{h_1}=0)\\
    - \frac{2\left[\log (m_\chi^2/m_{h_1}^2)+2\right]m_{h_1}^6}{v v_s^2}
    \sin^4\theta \cos\theta (v \sin\theta+v_s \cos\theta )
    & ({\rm for}~m_{h_2}=0)
  \end{array}\right. \,.
  \label{eq:self+vert|largemchi}
\end{align}

\subsection{Box and triangle diagrams}

\begin{figure}[t]
 \begin{center}
     \includegraphics[scale=0.65]{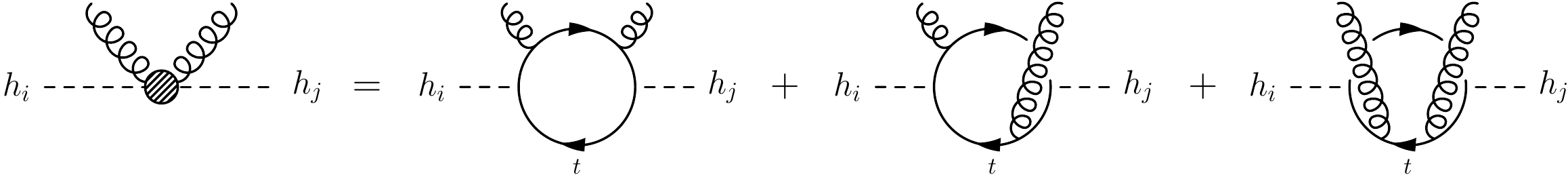}
  \caption{Higgs correlation functions.}
  \label{fig:HiggsGG}
 \end{center}
\end{figure}

To compute the Wilson coefficients induced by $\chi$--$g$ scattering,
it is legitimate to derive Higgs correlation functions shown in
Fig.\,\ref{fig:HiggsGG}.  We denote them as $\tilde{\Pi}_{ij}(q^2)$,
which are obtained straightforwardly by using the formula given in
refs.~\cite{Hisano:2010fy,Hisano:2010ct} as,
\begin{align}
  \tilde{\Pi}_{ij}(q^2)=c^G(q^2;\mu_W)
  \frac{\alpha_s}{\pi} G^a_{\mu\nu}G^{a\mu\nu}\,,
\end{align}
where
\begin{align}
  c^G(q^2;\mu_W)
  =-\frac{c_{qqh_i}c_{qqh_j}}{8m_\chi^2}I(-q^2/m_{\chi}^2,x_t)
  \,.
\end{align}
Here $I(t,x_t)$ is a dimensionless function,
\begin{align}
  I(t,x_t)=\frac{t-2x_t}{t(t+4x_t)}
  +\frac{2x_t(t+x_t)}{t^2(t+4x_t)\beta(t,x_t)}
  \log\left[\frac{2x_t+t(1+\beta(t,x_t))}{2x_t+t(1-\beta(t,x_t))}\right]\,,
\end{align}
with $\beta(t,x_t)=\sqrt{1+4x_t/t}$, $x_i=m_{h_i}^2/m_\chi^2$, and
$x_t=m_t^2/m_\chi^2$. Then, the Wilson coefficients coming from the box and
triangle diagrams are given by the following integrals,
\begin{align}
  &J_{\rm box}^{ij}=\frac{\kappa^{ij}}{8(4\pi)^2}\int^\infty_0 dt
  \frac{t I(t,x_t)}{(t+x_i)(t+x_j)}\left(1-\sqrt{(t+4)/t}\right) \,,\\
  &J_{\rm tri}^{ij}=\frac{1}{8(4\pi)^2}\int^\infty_0 dt
  \frac{t I(t,x_t)}{(t+x_i)(t+x_j)}\,,
\end{align}
with $\kappa^{ij}=2$ for $i\neq j$ otherwise 1. As mentioned in the
main text of the paper, $J^{ij}_{\rm tri}/m_\chi^2$ is constant with
respect to $m_\chi$. On the other hand, $J^{ij}_{\rm box}\propto
m_\chi^4$ ($m_\chi^3$) for small (large) $m_\chi$ region. Therefore
the contribution from the box diagrams is suppressed as $1/m_\chi$ in
large DM mass region while it becomes constant in small DM mass
region.

\appendix

\end{document}